\documentclass[12pt]{iopart}
\usepackage{graphicx}
\usepackage{epstopdf}
\usepackage{color}
\usepackage[usenames,dvipsnames]{xcolor}
\usepackage{umoline}
\usepackage{soul}

\newcommand{\ben}{\begin{eqnarray}}
\newcommand{\een}{\end{eqnarray}}

\newcommand{\be}{\begin{equation}}
\newcommand{\ee}{\end{equation}}
\newcommand{\ba}{\begin{eqnarray}}
\newcommand{\ea}{\end{eqnarray}}

\newcommand{\beq}{\begin{equation}}
\newcommand{\eeq}{\end{equation}}
\newcommand{\beqa}{\begin{eqnarray}}
\newcommand{\eeqa}{\end{eqnarray}}

\newcommand {\h}[1]{\hat#1{}}

\newcommand{\eqref}[1]{(\ref{#1})}

\begin{document}

\title[Emergent dark energy via decoherence in quantum interactions]{Emergent dark energy via decoherence in quantum interactions}

  \author{Natacha Altamirano$^{1,2*}$, Paulina Corona-Ugalde$^{1,3}$, Kiran~E.~Khosla$^{4,5}$, Gerard J. Milburn$^{4,5}$, Robert B. Mann${^1}$}
  \address{$^{1}$Perimeter Institute, 31 Caroline St. N. Waterloo
 Ontario, N2L 2Y5, Canada, \\
$^{2}$Department of Physics and Astronomy, University of Waterloo, Waterloo, Ontario, Canada, N2L 3G1, \\
$^{3}$Institute for Quantum Computing, University of Waterloo,
 Waterloo, Ontario, Canada, N2L 3G1, \\
$^{4}$Department of Physics, University of Queensland, St Lucia, QLD 4072, Australia, \\
$^{5}$Centre for Engineered Quantum Systems, University of Queensland, St Lucia, QLD 4072, Australia.}
\ead{$^*$naltamirano@perimeterinstitute.ca}

\vspace{10pt}
\begin{indented}
\item[]\today
\end{indented}

\begin{abstract}
In this work we consider a recent proposal in which gravitational interactions are mediated via classical information and apply it to a relativistic context. We study a toy model of a quantized Friedman-Robertson-Walker (FRW) universe with the assumption that any test particles must feel a classical metric. We show that such a model results in decoherence in the FRW state that manifests itself as a dark energy fluid that fills the spacetime. Analysis of the resulting fluid, shows the equation of state asymptotically oscillates around the value $w=-1/3$, regardless of the spatial curvature, which provides the bound between accelerating and decelerating expanding FRW cosmologies. Motivated by quantum-classical interactions this model is yet another example of theories with violation of energy-momentum conservation whose signature could have significant consequences for the observable universe.
\end{abstract}

\pacs{03.65.Ta  03.65.Yz  04.60.-m 98.80.Qc}
%\maketitle

\section{Introduction}\label{sec:1}

Since the beginning of the 20th century researchers have tried to understand the 
quantum description of the different interactions that describe nature. All 
forces in the standard model are currently understood in terms of local quantum interactions
and long range forces emerge as fluctuations of underlying gauge fields in the low energy limit, such as photons for the Coulomb force~\cite{baym_two-slit_2009}. Interactions in quantum field theory are described by quantum gauge fields that act as {\it{force carriers}} and, as they admit a quantum description, they can carry quantum information. Gravitation, however, remains stubbornly resistant to quantization.  The two most popular approaches, string theory \cite{Polchinski:1998rq} and loop quantum gravity \cite{Rovelli:2010bf}, have yet to attain their goals. Other approaches to  quantum gravity \cite{modesto_super-renormalizable_2012,stelle_renormalization_1977,moffat_finite_1990,cornish_quantum_1992,kiefer_conceptual_2013} suffer from non-local interactions or are non-renormalizable.

A number of authors have questioned if gravity needs to be quantized \cite{carlip_is_2008,albers_measurement_2008,boughn_nonquantum_2009} raising well known problems in consistently combining quantum and classical mechanics~\cite{peres_hybrid_2001,Ahmadzadegan:2016wsm}. In particular, gravitational decoherence models proposed by Diosi and Penrose \cite{diosi_models_1989,penrose_gravitys_1996} use principles of relativity to limit the lifetime of spatial quantum superpositions and, as a result, breaking the unitary evolution of the wavefunction.  One new approach along these lines \cite{2014NJPh...16f5020K} is the suggestion that gravity  is fundamentally classical and therefore cannot carry quantum information  \cite{nielsen_conditions_1999}. This approach is motivated by the fact that gravity cannot be shielded and therefore any observer can in principle gain information about  the quantum state sourcing gravity. The process of gaining (partial) information about a quantum state is equivalent to making weak measurements~\cite{aharonov_how_1988}, and is consistent with the standard approach for describing open quantum systems~\cite{gardiner_input_1985}. For example, a test particle in a quantum potential will become entangled with the source of the potential, and an observer who is not aware of the test particle (i.e. traces over the test particle degrees of freedom) will necessarily see decoherence in the evolution of the source particle. This decoherence mechanism is present for any quantum potential, for example the Coulomb interaction, and not limited to gravity. The distinction between the electric and gravitational potential is the ability to, in principle, shield this effect: a superconducting shell around a source charge eliminates the test-source interaction thereby shielding the decoherence; however
%\cyan{(I don't understand what is meant by "shielding the decoherence from the electric potential" should it be: "shielding the electric potential from decoherence?" )}
 there is no such shield for gravity. This form of decoherence,  perfectly compatible with the unitary evolution of standard quantum mechanics, motivated  consideration of a classical channel model for gravitational (CCG) interactions \cite{2014NJPh...16f5020K}. 

 In the CCG model, the gravitational potential is assumed to be fundamentally classical even though it can be sourced by quantum states. This quantum-classical interaction induces unavoidable decoherence on the quantum systems \cite{diosi_quantum_1995,diosi_quantum_2000}. This form of decoherence is not a consequence of tracing over an entangled state (as in the case of quantum potentials) but rather a  modification of unitary evolution as a consequence of quantum-classical interactions.  We will discuss the difference between CCG and standard unitary evolution in detail in Sec.~\ref{sec.modelcomparison}. 

Previous models of CCG have only considered Newtonian gravity, \cite{2014NJPh...16f5020K,kafri_bounds_2015,khosla_detecting_2016,Altamirano:2016fas}. In this paper we take the first steps towards a CCG model in the relativistic regime. The question we want to answer is how to apply the CCG model  when one has quantum metric degrees of freedom, and what are its consequences. We do this by considering a gravitational system with the fewest number of degrees of freedom possible, namely a canonically quantized empty Friedmann Robertson Walker (FRW) universe. In the Newtonian case, the trajectory of a test particle depends on the masses and configuration of the source, whereas in the GR description the dynamics are given solely by the metric components, i.e. the scale factor in an FRW spacetime. In CCG, the source necessarily experiences decoherence, and we will show that in the FRW context, CCG introduces decoherence of the spacetime. We will show that for an observer in such a universe this decoherence is  manifested as a time dependent dark fluid. \\

Our paper is organized as follows: In section~\ref{sec.modelcomparison} we compare and contrast the Newtonian formulation of CCG with our relativistic extension, and highlight the distinctions between decoherence predicted by CCG and the standard notion of decoherence obtained from unitary quantum mechanics. In section \ref{sec.cqmf} we will introduce the canonical spacetime Hamiltonian, and derive the equations of motion for the expectation values of the quantum observables. We analyze the solutions of the equations of motion in section \ref{sec.analysis} and compute the effective energy momentum tensor arising from this model. We then analyze the effective form of dark energy that emerges and consider how it affects the evolution of the spacetime.  We close with some final remarks  in section \ref{sec.conclusion} and discuss future directions for this approach. Through all this work we are considering units $G=1=c$.

\section{Relativistic Classical Channel Gravity}\label{sec.modelcomparison}
%\begin{itemize}
%%\item outline motivation for the section (mention de WdW)
%%\item explain in detail the Newtonian case-- difference btw unitary entanglement and CCG
%%\item explain Wheeler de Witt (unitary only)
%%\item euristic explanation of our model
%\item \tcr{Difference between non-unitary, decoherence}
%\end{itemize}

\begin{figure*}
\centering
  \includegraphics[width=1\linewidth]{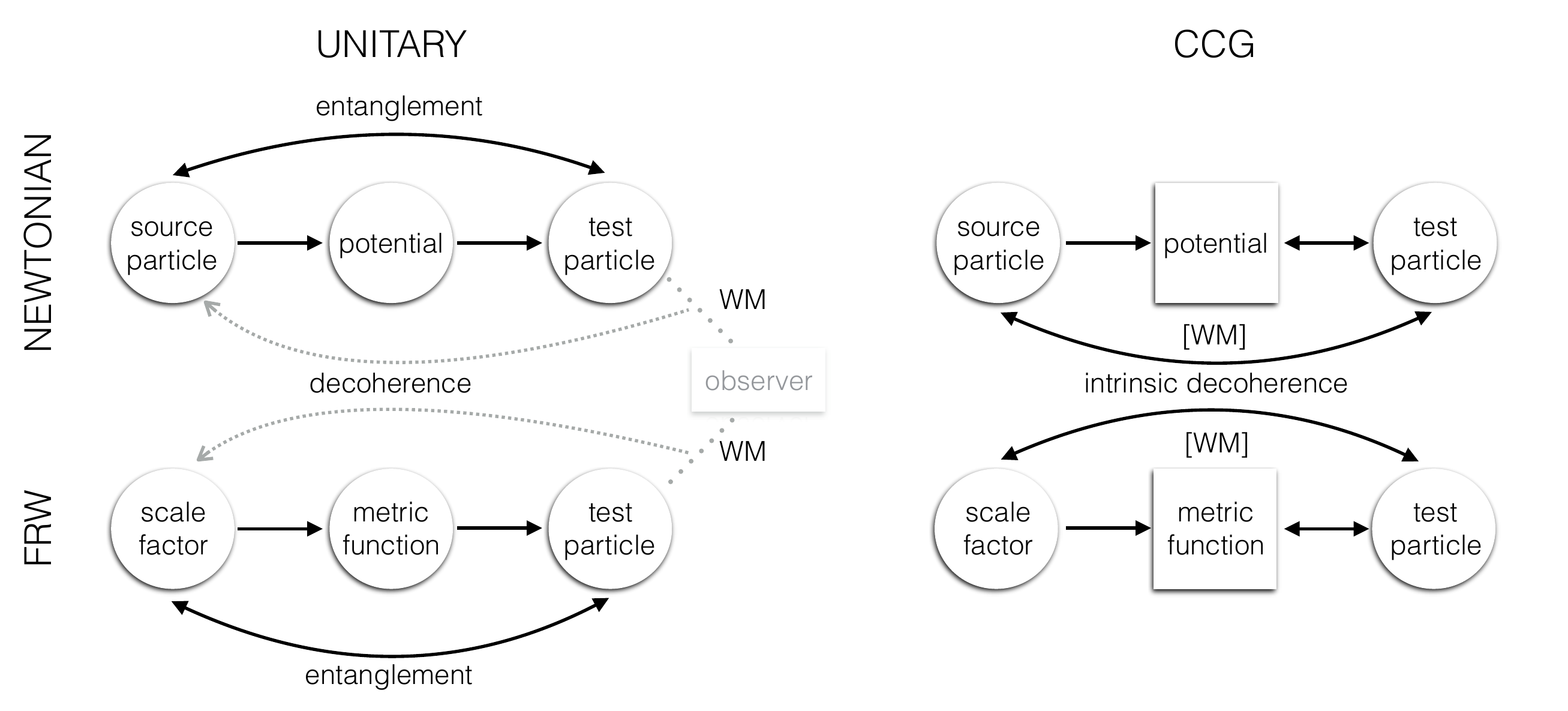}
 \caption{\label{fig:models}  GRAVITATIONAL MODELS. Cartoon of the four models presented in Sec.\ref{sec.modelcomparison}. We describe them by considering a source particle, a potential (top -- Newtonian) or metric (bottom -- cosmological) and a test particle reacting to the potential or metric. The circles represent quantum degrees of freedom whereas the squares represent classical degrees of freedom. For the unitary cases (left) the joint system source-potential (metric)-test particle evolve unitarily. In this case the source/test particle may become entangled and an observer making measurements on the test particle results in a weak measurement (WM) of the source, including the associated decohere. On the other hand, the CCG model (right) assumes that the only way a test particle can respond to the source is through classical information, which is mathematically equivalent to weak measurement and feedback control~\cite{2014NJPh...16f5020K}. This process results in decoherence for the source and classical fluctuations on the potential (metric) and therefore is fundamentally a non-unitary evolution for either quantum systems.  }
\end{figure*}

The goal of this section is to describe in detail the relativistic description of CCG. We do this by comparing unitary Newtonian gravity with the Wheeler de Witt equation, and show how the CCG model fundamentally differs from unitary dynamics. In particular we explain how the presence of a test particle in the Newtonian model of CCG results in decoherence of any object that sources a (gravitational) potential for that test particle. We then argue analogously that in relativistic CCG the presence of a test particle in an FRW universe leads to decoherence in the scale factor, and therefore non-unitary evolution of the universe.\\

%\tcb{by describing how standard unitary evolution of quantum systems under the Newtonian interaction is analogous to} the Wheeler de Witt equation. We then discuss how Newtonian CCG fundamentally differs from unitary dynamics and how this difference can be understood in the context of the relativistic description of CCG and how it compares to the Wheeler de Witt equation. 

{\it{Unitary Newtonian Interaction}}. Consider a quantum Newtonian interaction between a source and a test particle (Fig.~\ref{fig:models} top left). Under unitary evolution the two (perhaps distant) particles may become entangled, where such entanglement implicitly assumes a quantum ``force carrier" (potential) --- analogous to the photon in electrodynamics; i.e. if the source particle is in a quantum superposition, the test particle will feel a coherent superposition of potentials, and thus follow two trajectories in superposition. Any observer who makes a projective measurement of the position of the test particle is effectively making a weak measurement of the source particle, and this weak measurement induces decoherence of the source particle~\cite{aharonov_how_1988,jacobs_straightforward_2006}. This is an example of how two fundamental postulates of quantum mechanics (unitary evolution and the Born rule) lead to decoherence of a quantum state. In particular a projective measurement, i.e. an observer, is required for this type of decoherence and the fundamental evolution is unitary. 

{\it{Newtonian Classical Channel Gravity}}. The CCG model postulates that there is no quantum description for gravity and that the non-local interactions emerge from local interactions between a quantum particle and a classical potential. In this case, the gravitational interaction between a source and a test particle is mediated by a classical (as opposed to quantum) information channel (Fig.~\ref{fig:models} top right). The question is now how a quantum particle can source a classical potential and how a test particle responds to it. In Ref. \cite{2014NJPh...16f5020K} it was shown that such a classical information channel is equivalent to an agent performing weak measurements of the position of the source and using the measurement outcome to control a potential for the test particle. Consequently the test particle does not respond to a quantum potential generated by the source but rather to a classical estimate of this potential. In CCG the existence of a test particle responding to the potential necessarily results in decoherence and subsequent non-unitary evolution of the source, even in the absence of an observer making any measurements of the test particle.  
Finally, we note that this model goes beyond the standard postulates of quantum mechanics. We devote~\ref{app.ccgmathematics} to a mathematical description of CCG as presented in \cite{2014NJPh...16f5020K}.

The discussion so far has focused on the Newtonian description. Our goal is to understand how this same procedure can be carried out in a relativistic context. In the following we give a relativistic formulation of CCG in the cosmological context and compare it to the Wheeler-DeWitt equation. 

{\it{Wheeler-deWitt}}. The Wheeler-DeWitt equation, $\hat{H}|\psi\rangle = 0$ where $\hat{H}$ is the Hamiltonian operator of the spacetime including any matter and $|\psi\rangle$ is the quantum state of the universe, is the standard approach to quantum cosmology. In general relativity the least action principle always forces the classical Hamiltonian to vanish, and the Wheeler-DeWitt equation is the quantum implementation of this constraint.  We will restrict the following discussion to an empty FRW universe so the only spacetime observables are the scale factor $\hat{a}$ and its canonical conjugate momentum $\hat{\pi}$. If we now consider a quantum test particle moving in such a universe, we would expect the particle to become entangled with the state of the universe, exactly analogous to how a test particle becomes entangled with a source particle in unitary quantum mechanics (Fig.~\ref{fig:models} bottom left). By test particle we mean a particle whose contribution to the mass/energy of spacetime can be neglected, but is (in principle) able to become entangled with the spacetime state.
 In this context, we can view the scale factor as acting like a source that influences the dynamics of a test particle via the metric. Analogous to the unitary Newtonian case, in this scenario the presence of an observer making a measurement on the test particle is needed in order to get decoherence in the  quantum state of the universe, but othwerwise the evolution is entirely unitary. In the following we will use the analogy:  source $\to$ scale factor and  potential $\to$ metric to explain the main idea of this paper: the relativistic version of CCG. 

{\it{Relativistic Classical Channel Gravity}}. So far we have shown the interpretational similarity between unitary Newtonian gravity and the Wheeler-DeWitt equation:  both are mediated by a \textit{quantum potential} (metric). We are interested in understanding how CCG applies to the relativistic gravity: how a quantum spacetime  can influence the dynamics of a test particle via a classical metric  (Fig.~\ref{fig:models} bottom right).
Our interpretation in the empty FRW case is to view the quantum scale factor as a `source' of the classical metric, analogous to the way that a quantum particle sources a classical potential in  Newtonian CCG. A test particle in an FRW spacetime will follow a trajectory that solely depends on the scale factor, and thus the scale factor generates an effective potential for the test particle. Therefore, in analogy with the Newtonian CCG description, the scale factor-test particle interaction can be understood in terms of weak measurements and feedback control, and therefore there must be intrinsically non-unitary evolution of the quantum state.  

Using the same language of measurement and feedback from ref.~\cite{2014NJPh...16f5020K} we posit that the quantum state of the universe is subject to weak continuous measurement of the variable $\hat{a}^2$.  The measurement is of  $\hat{a}^2$, as opposed to $\hat{a}$, since classically it is the factor $a^2$ that appears in the metric function, and therefore the trajectory of any test particle can only depend explicitly on $a^2$. The measurement process forces the gravitational influence of spacetime  on the test particle to be mediated by classical information. In other words, the test particle responds to a classical estimate of the scale factor (the measurement results) analogous to the way that a test particle responds to the Newtonian potential in CCG as described in appendix A. 
%The dynamics of a single test particle in CCG will depend only on the results of an effective weak measurement, $\bar{a^2}$, where $\bar{a^2} = \langle \hat{a}^2\rangle_c + \sqrt{\frac{\hbar}{\gamma}}dW/dt$, the former quantity corresponding to a conditional measurement on  $\hat{a}^2$  and the latter quantity being the standard Wiener increment, as discussed in appendices A and B.    

The presence of the weak measurement on the quantum scale factor changes the evolution of the quantum state of the universe, resulting in a master equation for the ensemble averaged state that we shall describe in detail in the next section. An observer who tries to recover the dynamics of the scale factor, must measure the trajectories of many test particles and so cannot distinguish which sequence of measurement histories took place \cite{downes_optimal_2011,kar_maxwell_1993}; hence they observe an  ensemble averaged (i.e. averaged over all possible measurement histories) spacetime.
%is $\mathcal{E}(\bar{a^2})  = \langle \hat{a}^2(\tau) \rangle \equiv  \texttt{a}^2 $, where $\langle {A}\rangle=\mathrm{Tr}[A {\rho}]$ for a system with density matrix $\rho$. 

 We shall denote  $\texttt{a}^2$ as the classical scale factor experienced by any observer in the Universe. 
The relationship between $\texttt{a}^2$  and  $\hat{a}^2$ is described in~\ref{app.master} and in the following section. In our model, the evolution of $\texttt{a}$ is different from the standard Friedmann evolution and we will show that this is consistent with a dark energy fluid. 
Note that the effective measurement process avoids defining the classical scale factor as $\langle \h a\rangle^2$ or $T_{\mu\nu}=\langle \hat{T}_{\mu\nu} \rangle$ where the expectation value is calculated with the quantum state given by the Wheeler-DeWitt equation.

In this section we have described the relevant  properties of unitary evolution in both the Newtonian regime and in the cosmological scenario. These two models share the feature that interactions are mediated by quantum potentials that are able to entangle the interacting constituents.  On the other hand, CCG postulates that the gravitational interaction should be mediated by a classical potential, i.e. the interacting constituents (assumed to be quantum) will \textit{communicate} with each other by exchanging classical information. Such an interaction induces noise in the dynamics of the quantum constituents, and such noise can be modelled by a measurement feedback channel. The fundamental distinction between the unitary approach and CCG is studied in the next section for the cosmological case.

%%%%%%%%%%%%%%%%%%%%%%%%%%%%%%%%%%%%%%%%%%%%%%%%%%%%%%%%%%%%%%%%%%5
%%%%%%%%%%%%%%%%%%%%%%%%%%%%%%%%%%%%%%%%%%%%%%%%%%%%%%%%%%%%%%%%%55
\section{Model} \label{sec.cqmf}

In this section we present the details of cosmological CCG. We begin with a classical FRW metric describing an empty, isotropic, homogenous universe. In conformal time, the line element is 
\be
 ds^2=a^2(\tau)[-N(\tau)^2d\tau^2+\frac{1}{1-kr^2}dr^2+r^2d\Omega^2]\,,
 \label{FRWmetric}
\ee
where $a(\tau)$ and $N(\tau)$ are the time dependent scale factor and lapse function respectively. The curvature term $k=-1,0,1$ describes open, flat, and closed spatial slices of the universe respectively. The Einstein-Hilbert action for an empty spacetime is
\be\label{einact}
S=\frac{1}{16 \pi}\int d^4x \sqrt{-g}R\,,
\ee
where $g$ is the determinant of the metric $g_{\mu\nu}$ and $R$ is the Ricci scalar. From the action, the Lagrangian and Hamiltonian density can be found to be
 	\begin{eqnarray}
	{\cal{L}}&=&k a^2 N-\frac{\dot{a}^2}{N},\label{efflag} \\
	%H &=& \int N{\cal{H}}dt \qquad 
	{\cal{H}} &=& -\frac{\pi^2}{4}-ka^2\,,\,\,\,\,\,\,\,\,\,\,\,\,\,
	\label{eq:hamilt}
	\end{eqnarray}
where $\pi={-2\dot{a}}/{N}$ is the canonical momentum  conjugate to $a$, with dots denoting conformal time derivatives. Note that the absence of any spatial dependence in \eqref{eq:hamilt} means the  Hamiltonian density proportional to the full Hamiltonian, and so we shall take ${\cal{H}}$ to be the Hamiltonian (up to a factor with units of volume). The lapse function $N(\tau)$ acts as a Lagrange multiplier, ensuring the classical constraint ${\cal{H}}=0$ and it can be chosen to be unity without loss of generality. The quadratic form of the Hamiltonian density in~\eqref{eq:hamilt} is due to the choice of conformal time in the definition of the line element in equation~\eqref{FRWmetric}. Hamilton's equations of motion are
\begin{eqnarray}
\dot{a} &=& -\pi/2 \,,\label{eq:cl1}\\
\dot{\pi} &=& 2 k a.
\label{eq:cl2}
\end{eqnarray}

In the standard approach to quantum cosmology, the Hamiltonain becomes a quantum degree of freedom
${\cal H}={\cal H}(\hat{a}, \hat{p})$, and along with the state of the universe $|\psi(a)\rangle$ are required to obey
the  Wheeler-DeWitt equation
\begin{eqnarray}
\hat{{\cal{H}}}|\psi(a)\rangle =0 \Rightarrow
\frac{\partial^2 \psi}{\partial a^2} - k a^2 \psi=0\,,
\label{wheeler}
\end{eqnarray}
which is the quantum implementation of the Hamiltonian constraint\footnote{If matter degrees of freedom are introduced they can be modelled by introducing an additional term proportional to $a^3\mu\psi$ on the left-hand side of \eqref{wheeler}, 
where $\mu$ can be either a matter field or a phenomenological perfect fluid.}.  The Wheeler De-Witt equation, \eqref{wheeler}, can also be written in terms of the von-Neumann equation for the quantum density matrix $\hat{\rho}=|\psi(a)\rangle\langle\psi(a)|$ 
\be\label{wheeler2}
\frac{d\hat{\rho}}{d\tau} =  -\frac{i}{\hbar}[\hat{\cal H},\hat{\rho}] = 0\,,
\ee
where  $\tau$ is the time flow used to define the operator $\pi$ in the Legendre transformation. Different interpretations of the meaning of $|\psi\rangle$ lead either to single patch models (where the whole universe is thought of as a collection of interacting homogeneous patches) or to minisuperspace models (where the patch is the whole universe at early times when no inhomogeneities had formed).  Either approach leads to the well-known problems of time and interpretation of the wavefunction in
 quantum cosmology  \cite{Wiltshire}, and efforts to solve them have led to a range of different  models  \cite{ MONIZQC, Calcagni10, Maartens2010, tachyon}.
In either case one then is left with the problem of both computing the wavefunction (or density matrix) of the universe and of interpreting it in such a manner that admits a reasonable classical limit.  

Here we seek to avoid the complication of a universal wave function and its associated interpretive issues by considering the relativistic extension of CCG  in which the fundamental degrees of freedom of the spacetime  remain quantum, but particles can only react to classical estimates of the underlying quantum observables.   In the FRW context this implies that quantum scale factor acts as source of the metric potential whose evolution (by the postulate of CCG) cannot become entangled with any matter, such as a test particle. An observer, wanting to describe the spacetime dynamics will make measurements on many test particles and thus the metric this observer perceives is 
\begin{eqnarray}
ds^2 = \texttt{a}^2(\tau)[-d\tau^2+\frac{1}{1-kr^2}dr^2+r^2d\Omega^2] \,,
\label{eq:Ds2}
\end{eqnarray} 
where we  have set  $N=1$ and $\texttt{a}^2 \equiv \langle \hat{a}^2\rangle = \mathcal{E}(\bar{a}^2) $ is the  classical information (estimate) gained from the quantum state of the spacetime experienced by each test particle (described in~\ref{app.master}). Note that  while  the scale factor and its canonical momentum are quantized, all gravitational effects are governed by $\texttt{a}$.  At early times (as we shall see) this avoids a singular universe due to the uncertainty principle. 

The gain in classical information of $\texttt{a}^2$ implies that  eq. \eqref{wheeler2} is no longer applicable since any gain in classical information about a quantum state, necessarily perturbs the state (see~\ref{app.master}). Since $\hat{a}^2$ is the observable  from which $\texttt{a}^2$ is estimated, we posit that  
\begin{eqnarray}
\frac{d\hat{\rho}}{d\tau}  &=& -\frac{i}{\hbar}[\hat{\cal H},\hat{\rho}] -\frac{\gamma}{8\hbar}[\hat{a}^2,[\hat{a}^2,\hat{\rho}]]\,,
\label{eq:Uncond}
\end{eqnarray}
governs the quantum evolution of the universe instead of \eqref{wheeler2}.  In contrast to the Wheeler De-Witt equation, which  suggests $[\mathcal{H}, \hat{\rho}] = 0$, in CCG the presence of the decoherence term in general perturbs the state from a Hamiltonian eigenstate, affecting the dynamics of the quantum system.

 The latter term in  \eqref{eq:Uncond} takes into account both the non-unitary evolution of the scale factor due to the presence of  test particle(s) and the ensemble average of an observer when making measurements on multiple test particles. It can be derived from a collisional model in which the quantum degrees of freedom are continuously interacting with the external test particles. This process introduces a new fundamental constant $\gamma$ that emerges as a consequence of the interaction between the test particle and the scale factor via the metric function. We give a full mathematical derivation of this equation in the~\ref{app.master}, and also show that the uncertainty principle and positivity of $\hat{\rho}$ holds at all times provided  $\gamma > 0$  \cite{lindblad_generators_1976,wiseman2009quantum}. The form of equation \eqref{eq:Uncond} have been previously used in different scenarios, in particular to account for modifications of quantum mechanics such as collapse models~\cite{bassi_models_2013} and with a focus on cosmological consequences of metric theories with non conservation of energy momentum tensor \cite{PhysRevLett.118.021102}. Here we go a step further, and analyze the consequences in the cosmological scenario of \eqref{eq:Uncond} when emergent form Quantum-classical interactions as the master equation for observables measured by a classical observer.

We emphasize that  the interpretation of this model is  fundamentally different from that of the standard Wheeler-DeWitt approach in \eqref{wheeler2}. Here the evolution of the universe is obtained by solving the master equation~\eqref{eq:Uncond}, for the time dependence of  $\texttt{a}^2 =   \langle \hat{a}^2(\tau) \rangle$
as in equation \eqref{eq:q-metric}.
 The resulting spacetime will behave very differently compared to that of an empty universe, particularly at early times. As we shall demonstrate, the universe described by \eqref{eq:Ds2} will evolve as though there were a form of time-dependent dark energy present.
  
The evolution of $\texttt{a}^2$ is solved by using the master equation \eqref{eq:Uncond} and computing time derivatives of the first and second order moments of the quantum operators $(\hat{a},\hat{\pi})$. In particular, we note that by construction that $d\,\texttt{a}^2/d\tau = \mathrm{Tr}[\hat{a}^2 d\hat{\rho}/d\tau]$ and thus we need to solve the following coupled equations 
\ben
~~~~~~~~\frac{d}{d\tau}{\langle \hat{a} \rangle}&=&-\langle \hat{\pi}\rangle/2\,, \label{A1} \\
~~~~~~~~\frac{d}{d\tau}{\langle {\hat{\pi}} \rangle}&=& 2k \langle \hat{a}\rangle\,,  \label{Pi2} \\
~~~~~~~~\frac{d}{d\tau}{\langle \hat{a}^2 \rangle}&=& -\langle \hat{a}\hat{\pi}+\hat{\pi}\hat{a}\rangle/2\,, \label{A2}\\
~~~~~~~~\frac{d}{d\tau}{\langle \hat{\pi}^2 \rangle}&=& 2k\langle \hat{a}\hat{\pi}+\hat{\pi}\hat{a}\rangle + \gamma \langle \hat{a}^2 \rangle\,, \label{Pi2}\\
\frac{d}{d\tau}{\langle \hat{a}\hat{\pi}+\hat{\pi}\hat{a}\rangle}&=& -\langle\hat{\pi}^2\rangle +4k\langle \hat{a}^2\rangle\,, \label{Comb2}
\een
where the time derivative of an expectation value $\langle \hat{X}\rangle$ is given by $d \langle \hat{X} \rangle /d\tau = \mathrm{Tr}[\hat{X} d\hat{\rho}/d\tau]$ for any operator $\hat{X}$. We see from equations \eqref{A1}--\eqref{Comb2} that only the second order moments are required to obtain the evolution of \eqref{eq:Ds2}.  Solving for these yields the evolution of the spacetime metric, and our subsequent task is to solve this set of equations for a variety of initial conditions for different values of the curvature constant $k$.

We find there is  always one exponentially growing mode, which makes the universe expand
and two modes that either yield exponential decay or decaying oscillation of the scale factor.  A general solution is a linear combination of these eigen-solutions, and so will in general asymptote to one that grows exponentially with time.

We now proceed to interpret these solutions from the perspective of an observer who only has access to to the trajectories of test particles to back out the metric \eqref{eq:Ds2}.

\section{Dark Energy from Decoherence}\label{sec.analysis} 

As noted above, an observer can in principle determine the temporal evolution of the observable $\texttt{a}^2$. Having no direct access to the underlying quantum observables, this observer can compute the Einstein tensor associated with the metric \eqref{eq:Ds2} and then use 
 Einstein's equations to determine the effective stress-energy tensor governing the observed evolution of spacetime.  

The solution  for  $\texttt{a}^2(\tau)$  depends on the six variables $\{\tau,k,\gamma,\texttt{a}^{(2)}_0,\pi^{(2)}_0,\zeta_{0}\}$, where $\tau$
is the conformal time 
%$k$ the spatial curvature, already defined
and  $\{\texttt{a}^{(2)}_0,p^{(2)}_0,\zeta_{0}\}$ are the second order moments of the quantum state $\{\langle \hat{a}^2\rangle, \langle \hat{p}^2 \rangle, \langle \hat{a}\hat{p} + \hat{p}\hat{a}\rangle\}$ at $\tau=0$. These quantities can be constrained using a variety of physical criteria, as we shall discuss in the next section.   

\begin{figure}
\centering
 \includegraphics[width=0.5\linewidth]{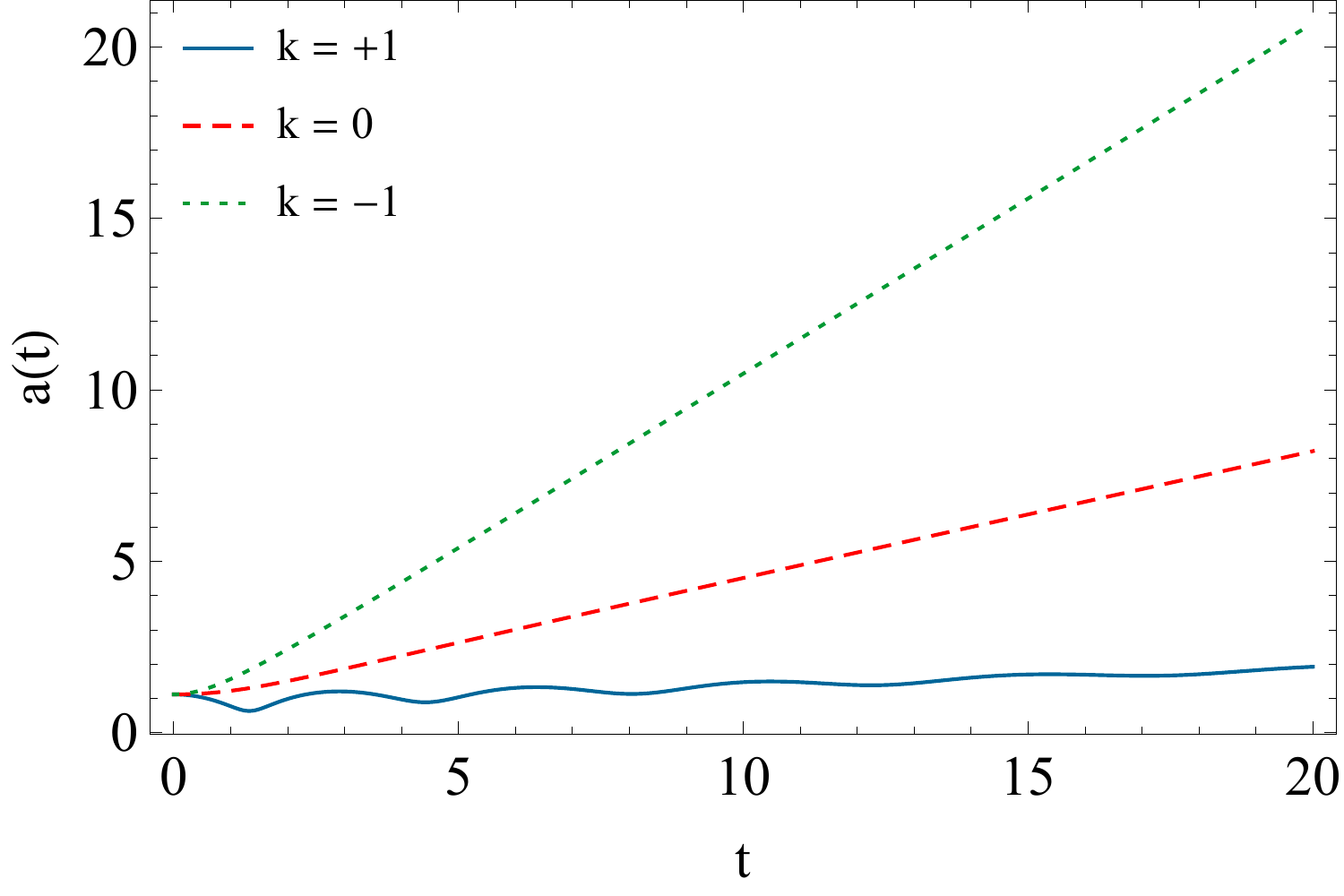}
 \caption{\label{fig:acomparison}Scale factor behaviour for $\gamma = 0.1$ and different
 values of $k$ as a function of comoving time. The quantum is system initially in a coherent state centered at $a_0 = 1$. The relative behavior remains qualitatively the same as the decoherence parameter $\gamma$ is varied.}
\end{figure}

To see the general dependence of  $\texttt{a}^2$ on $\gamma$, we set $\texttt{a}^{(2)}_0=1+1/4,~\pi^{(2)}_0 = 1~,\zeta_0 = 0$, describing displaced ground state of the quantum state and plot the results in comoving time in Fig.~\ref{fig:acomparison} for each value of $k$.  To   better understand the dynamics of the scale factor, and indeed the general physics of our model   (anticipating a comparison with data), it is useful to analyze relevant physical quantities in terms of the comoving time  $t$
\beq
t (\tau)=\int_0^\tau \texttt{a}(\tau')d\tau' \,,
\eeq
with $\tau$  being the conformal time. 
We see that in the case $k>0$ for small $\gamma$ the scale factor undergoes damped oscillations. Furthermore, there exists a time scale that is half the e-folding time associated with the positive root $\lambda_+$, after which the scale factor grows linearly in comoving time
(exponentially in conformal time) without oscillations as noted above. This growth, while present, is not visible for the most oscillatory curve in figure~\ref{fig:acomparison}. For the $k\leq 0$ cases there is exponential growth but no oscillations for this choice of parameters. We also note that the growth of the scale factor is faster for $k>0$ and slower for $k=0$.

As described in the introduction, from the point of view of an observers will infer from the motion
of test particles the metric \eqref{eq:Ds2}, and describe the resultant spacetime dynamics with  the Einstein equations
 \be 
 G_{\mu\nu}(\texttt{a}^2) =8{\pi}T_{\mu\nu}\,,
 \label{einseq}
 \ee
 where $G_{\mu\nu}=R_{\mu\nu}-\frac{1}{2}Rg_{\mu\nu}$ is the Einstein tensor. The Ricci tensor 
 $R_{\mu\nu}$ and  Ricci scalar $R$ are constructed with second derivatives of the metric tensor $g_{\mu\nu}$ which is given by \eqref{eq:Ds2}.  
 Such an observer  will infer that the expansion is driven by a form of {\it{dark energy}}, whose effective stress-energy is $T_{\mu\nu}$, and which we shall now  compute. 
 
 The symmetries of the FRW metric (homogeneity and isotropy) tell us that the energy-momentum tensor
 must have the form of a perfect fluid
 \be\label{pfluid}
 T_{\mu}^{\nu}=
\left(\begin{array}{cccc}
 -\rho & 0 & 0 & 0 \\
 0 & P & 0 & 0 \\
 0 & 0 & P & 0 \\
 0 & 0 & 0 & P \\
\end{array}
\right)\,,
\ee
where $\rho$ and $P$ are the energy density and pressure of our perfect fluid. The type of matter is characterized by $w$ in the equation of state $P = w \rho$. 

\begin{figure*}
\minipage{0.3\textwidth}
  \includegraphics[width=\linewidth]{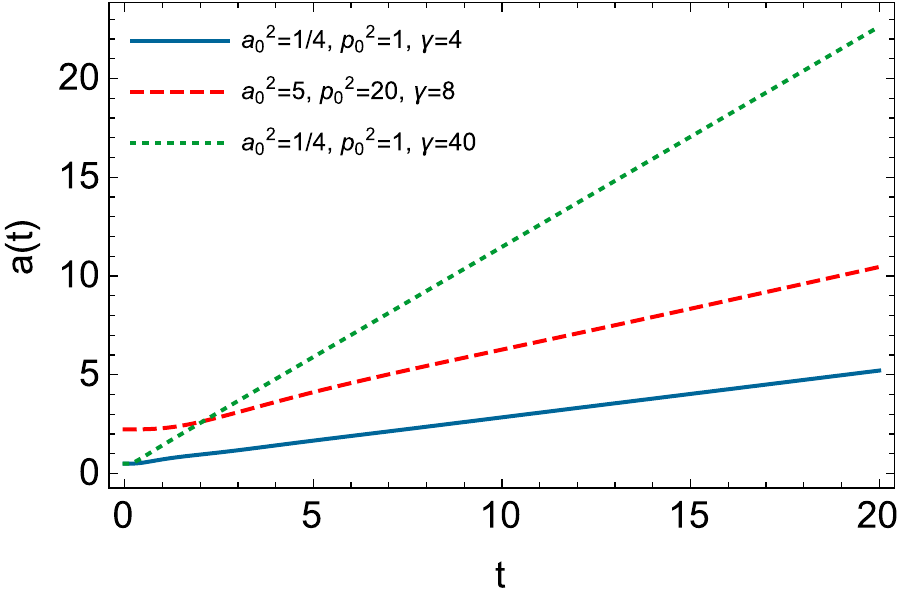}
  \endminipage\hfill
\minipage{0.32\textwidth}
  \includegraphics[width=\linewidth]{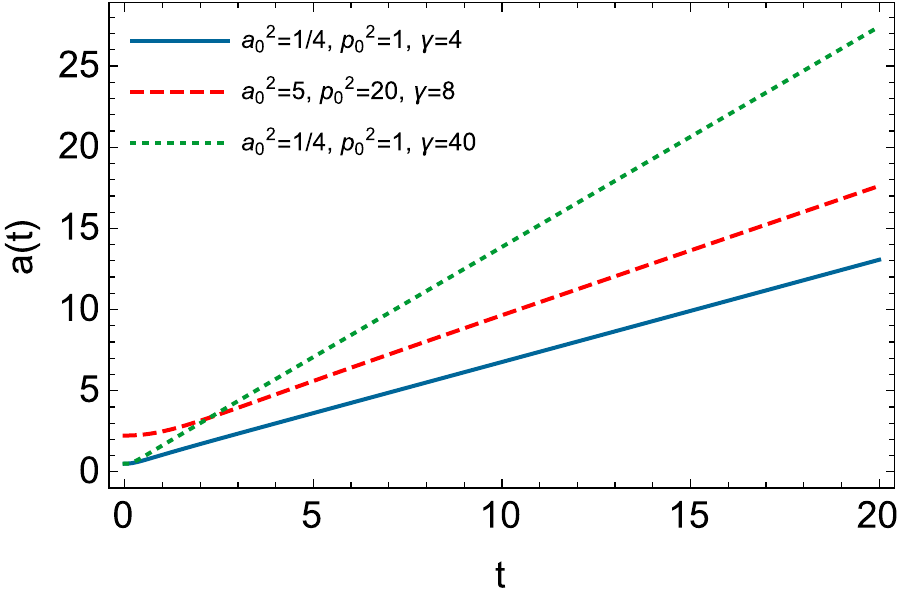}
  \endminipage\hfill
\minipage{0.32\textwidth}
  \includegraphics[width=\linewidth]{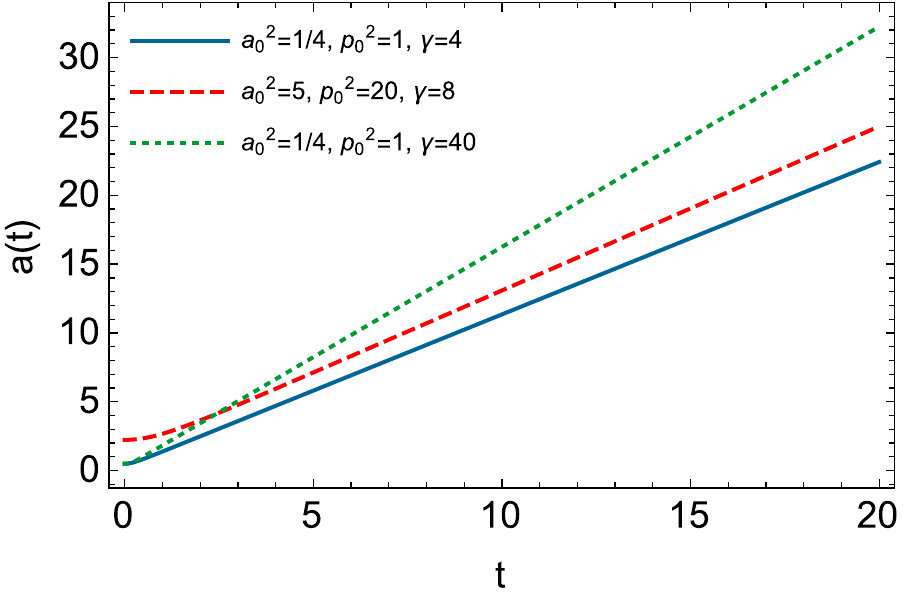}
  \endminipage\\
\minipage{0.3\textwidth}
  \includegraphics[width=\linewidth]{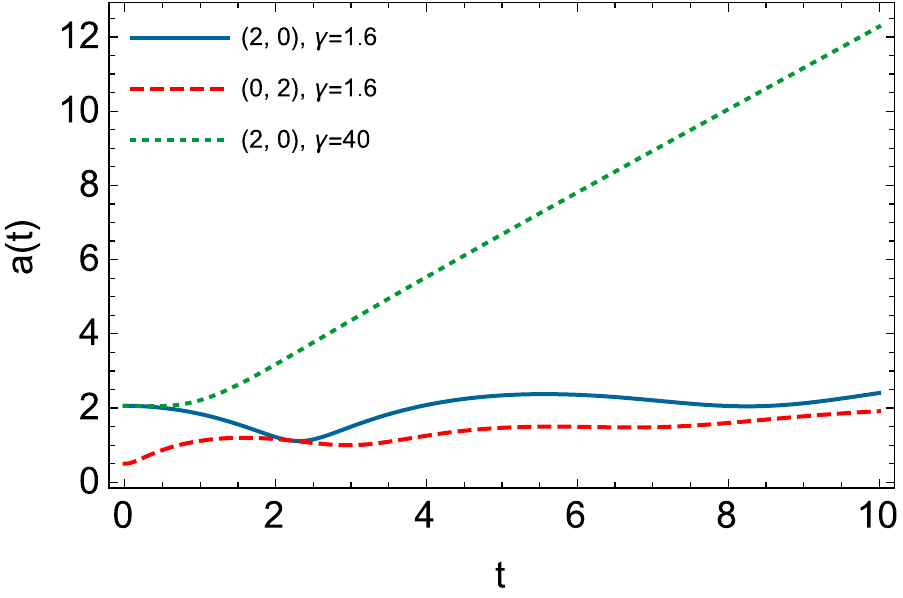}
  \endminipage\hfill
\minipage{0.32\textwidth}
  \includegraphics[width=\linewidth]{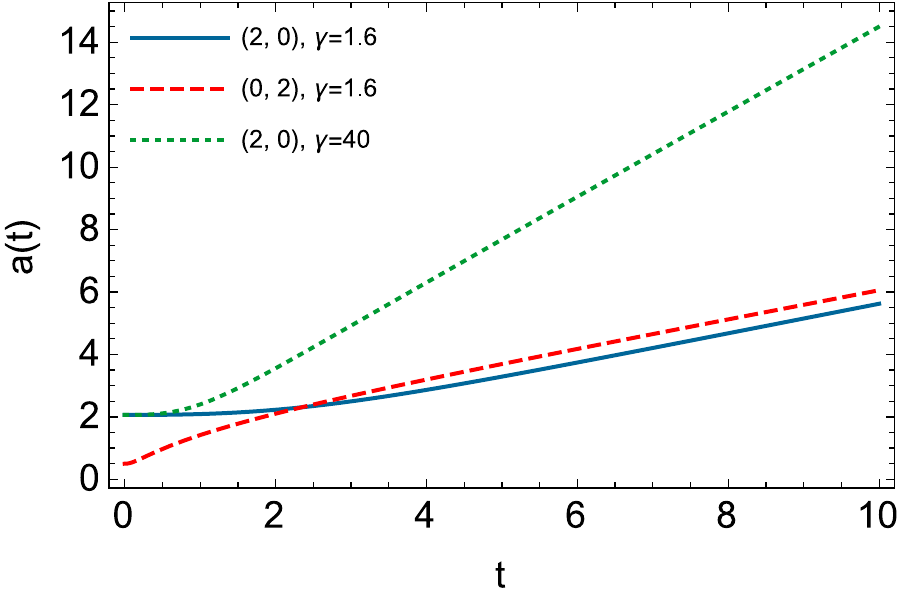}
  \endminipage\hfill
\minipage{0.32\textwidth}
  \includegraphics[width=\linewidth]{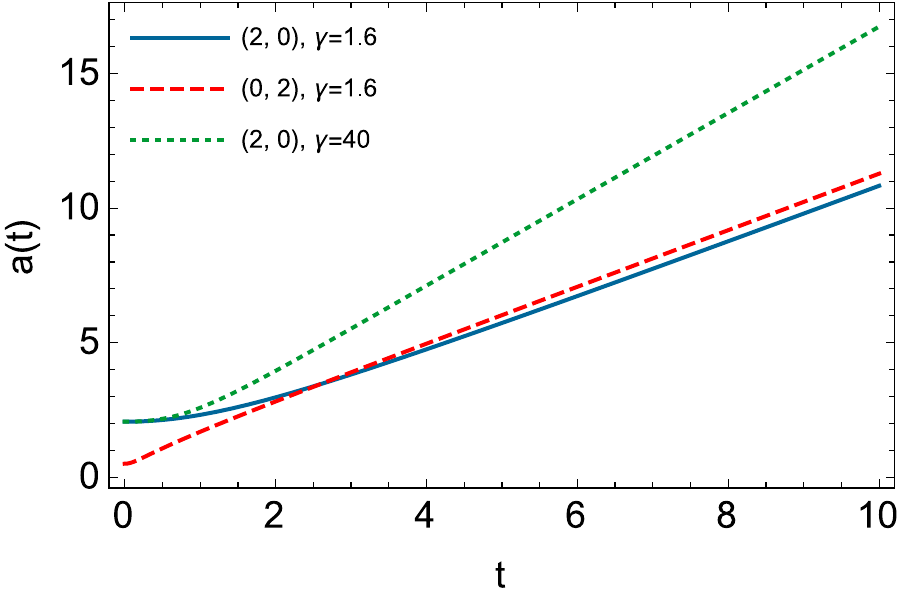}
  \endminipage
  \caption{\label{fig:a0}  Evolution of the scale factor in comoving time for initial thermal (top) and coherent (bottom) states, for $k = 1$ (left), $k = 0$ (middle) and $k = -1$ (right). In each case the steady state gradient is determined by $\gamma$. The oscillation in the $k=1$ case can be understood by considering the equations of motion (eq~\ref{a2}-\ref{comb2}) as $\gamma \rightarrow 0$. The equations are then that of a Harmonic oscillator, with the amplitude of the oscillations  defined by the Euclidean sum of the coherent amplitude $(a_0, p_0)$, where $\tan^{-1}(a_0/p_0)$ defines the initial phase.}
 \label{a2vtkp}
\end{figure*}

We must also choose the free parameters $\{a_0,p_0,\texttt{a}^{(2)}_0,\pi^{(2)}_0,\zeta_0,\gamma\}$ where  ${a}_0$ and ${\pi}_0$ are the initial means of $\hat{a} $ and $\hat{p}$ respectively. This is a \linebreak rather large parameter space, and constraining it to obtain useful information is a bit of a challenge.  We will consider two kinds of Gaussian states: coherent states saturating the uncertainty inequality, characterized by their mean amplitude $(a_0, p_0)$ (with $\zeta_0 = a_0\pi_0$), and thermal states for which $4a^{(2)}_0 = \pi^{(2)}_0$ with $\zeta_0 = a_0 = \pi_0 = 0$. The factor of four difference comes from the transformation of equation~(\ref{eq:hamilt}) into natural units. Note that the spacetime only depends on $\texttt{a}^2 =  \langle{\hat{a}^2}\rangle$, which in turn is governed by equations~\eqref{a2}-\eqref{comb2}. Consequently, initial values of $a_0 = \pi_0 = 0$ result in a spacetime driven by noise from either a quantum (for minimum uncertainty states), or quantum and statistical (for mixed states) source. Additionally we can impose the following physical conditions on the spacetime

\begin{itemize}   
%\item What type of fluid is $w(t)$ describing?, Baryonic, Dark Matter, Dark energy, radiation?,  
    \item Strong Energy Condition: $\rho>0$ and $w(t)>-1$\,.
    \item Weak Energy Condition: $\rho>0$ and $|w(t)|<1$\,.
		\item A non-singular spacetime:  $K<\infty$, where $K$ is the Kretschmann scalar $K=R^{abcd}R_{abcd}$. 
\end{itemize} 
Finally, a more sophisticated model would have to be observationally constrained by early-universe data from the CMB, as well as from information on structure formation, but this is beyond the scope of this toy model.

We have not imposed the Hamiltonian constraint  ${\cal{H}}=0$, required in  the standard picture of classical cosmology. The noise induced by the test particles measurements will break unitary evolution (as given by the Wheeler-DeWitt equation) as discuss in Sec.\ref{sec.cqmf}; consequently the spacetime Hamiltonian \eqref{eq:hamilt} becomes time dependent
\beq
\frac{d\cal{H}}{d\tau}=-\frac{\gamma}{4} \texttt{a}^2(\tau) \,.
\eeq
As outlined in section~\ref{sec:1}, it is unsurprising that the spacetime Hamiltonian alone is not conserved; indeed it is exactly this energy that gives rise to the gravitational source whose effective stress-energy tensor is given by \eqref{pfluid}. Note that the non conservation of the Hamiltonian of an empty universe, is governed by the decoherence parameter $\gamma$, and thus is deeply connected to the noise introduced by the test particles. Despite the  metric being purely classical, its components are affected by the
intrinsic quantum noise introduced by the interaction with test particle. When we compute the Hamiltonian using $\texttt{a}^2$ and its conjugate momentum we thus find there is an excess of energy in comparison with the classical case.

We now concentrate in the description of the perfect fluid and an analysis of singularities.  We  find that the Kretschmann scalar
\beq
K=\frac{12}{\texttt{a}^4}\bigg[(k+\texttt{a}'^2)^2+\texttt{a}^2\texttt{a}''^2\bigg]\,,
\eeq
does not diverge, since we must have $\texttt{a}>0$ to have a physical quantum state. 
The expression for the effective density is
\beq
\rho(t)=\frac{3}{8\pi}\frac{\texttt{a}'^2+k}{\texttt{a}^2}\,,
\eeq
and the equation of state is described by the function $w(t)$, defined via
\beq
P(t)=w(t)\rho(t)\,,
\eeq
where we find
\beq
w(t)= - \frac{2}{3}\frac{\texttt{a}''\texttt{a}}{\texttt{a}'^2+k}-\frac{1}{3}\,,
\label{eqofstate}
\eeq
where we have used Einstein equations~\eqref{einseq} for  $\texttt{a}^2$ assuming a perfect fluid~\eqref{pfluid}, and where $a'=\frac{da}{dt}$.

\begin{figure}
\centering
 \includegraphics[width=0.5\linewidth]{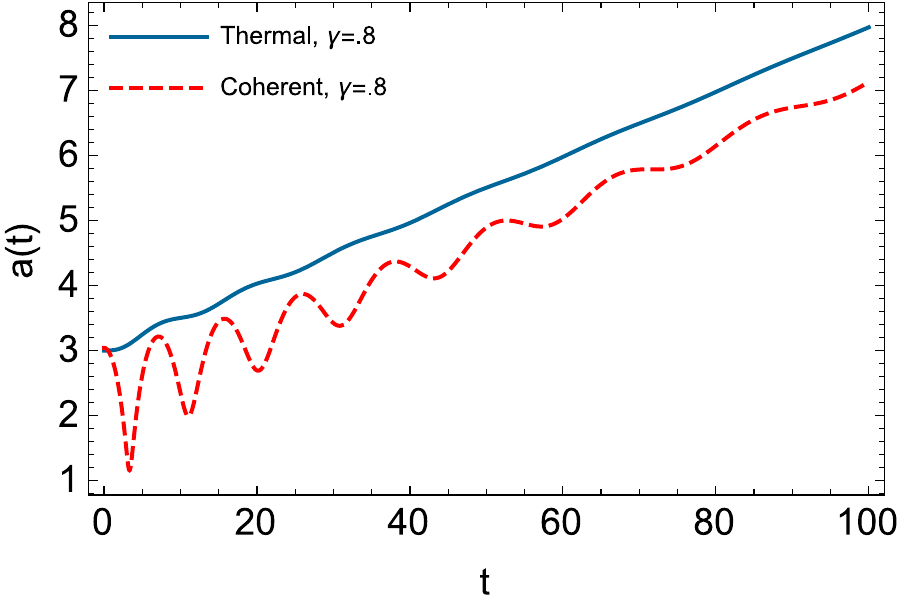}\\
 \caption{Evolution of the scale factor for $k = 1$. The large initial amplitude causes many oscillations before the transient behavior becomes dominated by the decoherence. The initial coherent state begins with an amplitude of $(3, 0)$ and is compared to a thermal state of $a_0^{(2)} = 9$. Without decoherence, the thermal state would remain constant in time.}
 \label{fig:Amplitude}
\end{figure}

We plot the behavior of the scale factor as a function of comoving time for different parameters. The results are depicted in figure \ref{a2vtkp}.  We see that the cosmic evolution has low sensitivity to the initial conditions indeed at late times, the different curves become indistinguishable for all values of $k$.  However there is rather high sensitivity to the choice of $\gamma$, particularly at early times, as is clear from figure  \ref{a2vtkp}. For small $\gamma$ the coherent behavior is resolvable for a longer time. This is clearly visible in the case of a large coherent amplitude, shown in figure~\ref{fig:Amplitude}, where the large initial coherent amplitude causes many visible oscillations before the spacetime is dominated by the decoherence. 

The behaviour of the Hubble parameter  $H= \frac{\texttt{a}'}{\texttt{a}}$ 
as a function of comoving time  (not plotted) displays  considerable sensitivity  for various choices of $\pi$ and $\gamma$ at early times, but  convergence at late times.   We also find that the   effective energy density $\rho$
%we see from  figures \ref{rhovtkp} and \ref{rhovtkg}
is always positive, and that at early times its behavior can be quite oscillatory for small values $\gamma$ when $k=+1$, but for all values of $k$ it tends to grow initially for various choices of $\pi$ and $\gamma$. At late times, regardless of these choices and values of $k$, the energy density is a monotonically decreasing function of time.

\begin{figure*}
\minipage{0.3\textwidth}
  \includegraphics[width=\linewidth]{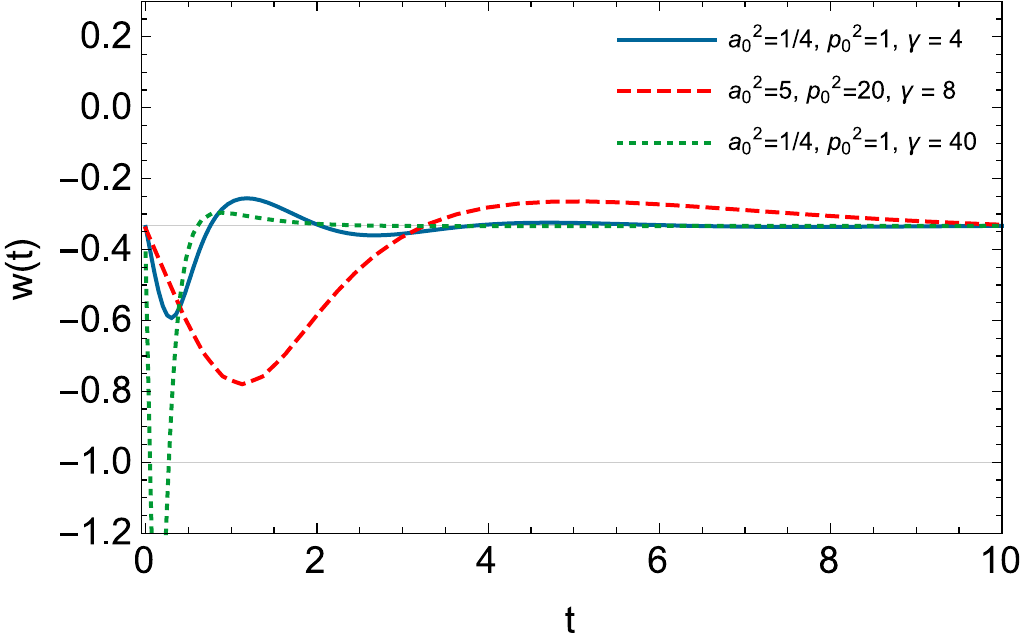}
  \endminipage\hfill
\minipage{0.32\textwidth}
  \includegraphics[width=\linewidth]{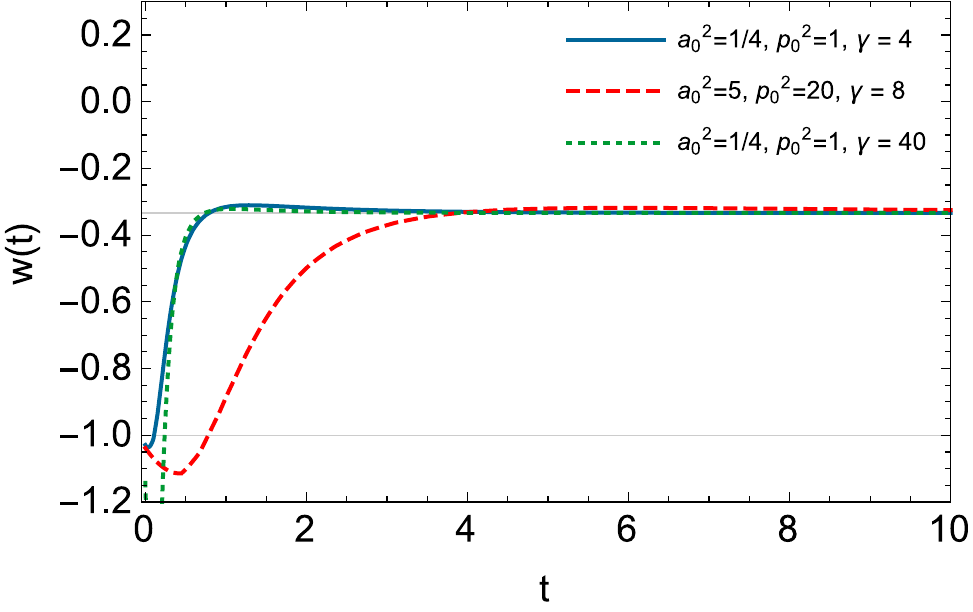}
  \endminipage\hfill
\minipage{0.32\textwidth}
  \includegraphics[width=\linewidth]{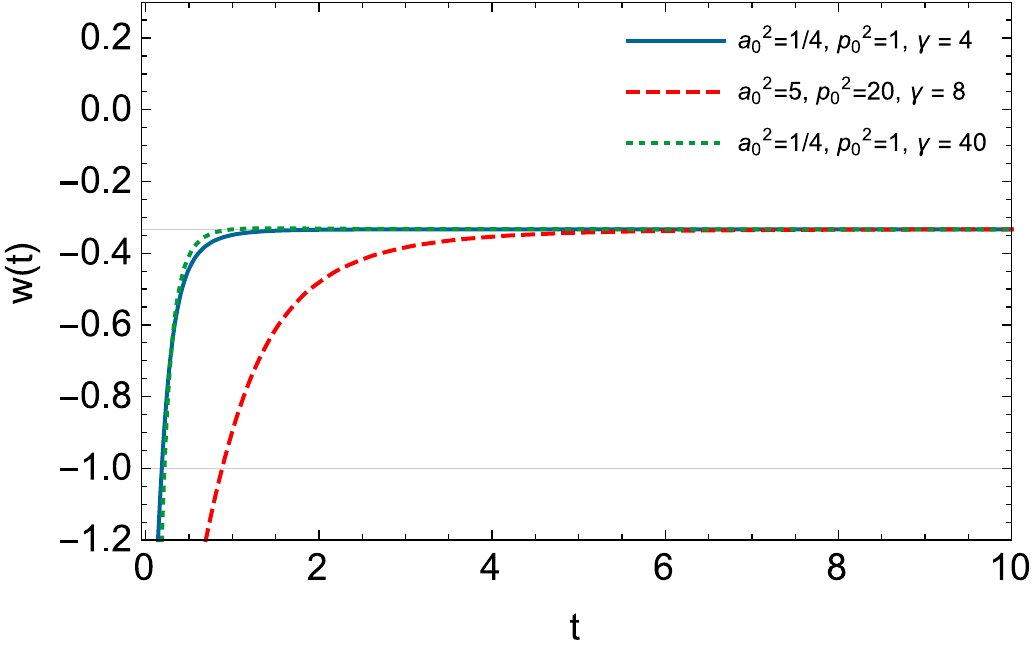}
  \endminipage\\
\minipage{0.3\textwidth}
  \includegraphics[width=\linewidth]{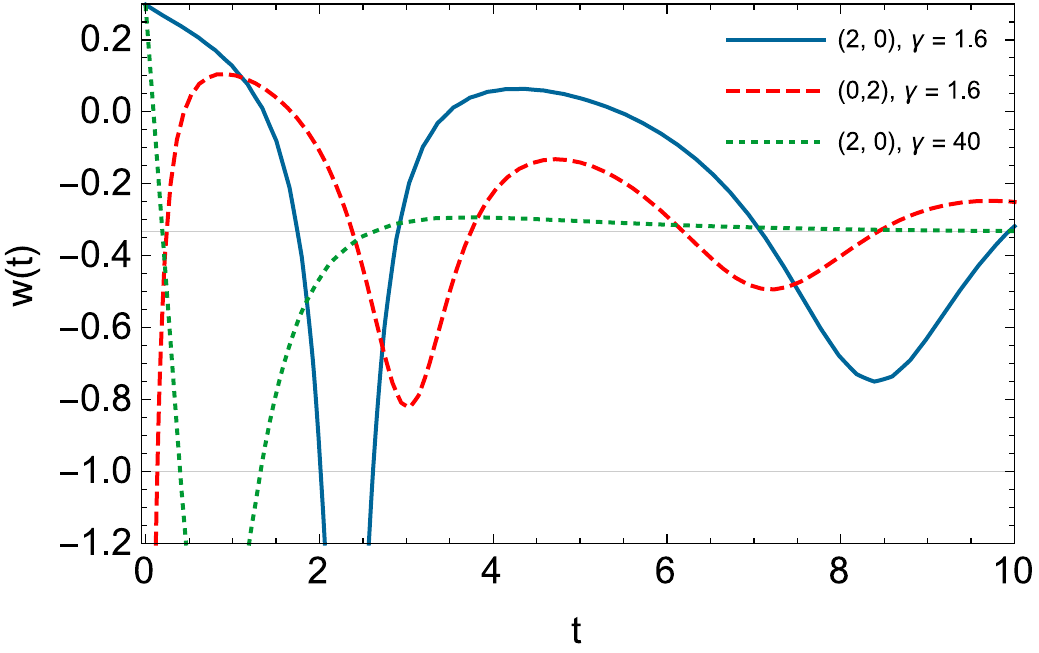}
  \endminipage\hfill
\minipage{0.32\textwidth}
  \includegraphics[width=\linewidth]{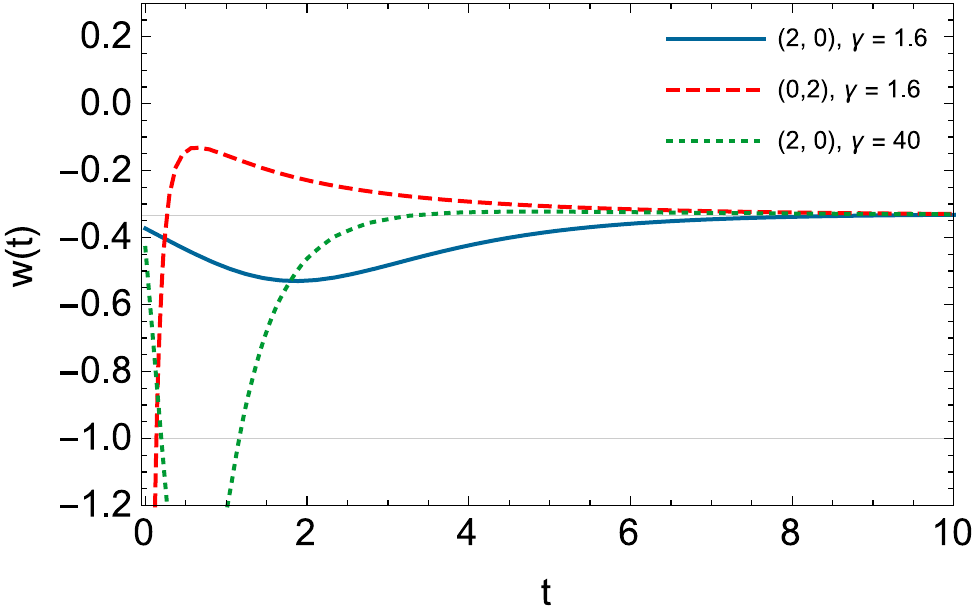}
  \endminipage\hfill
\minipage{0.32\textwidth}
  \includegraphics[width=\linewidth]{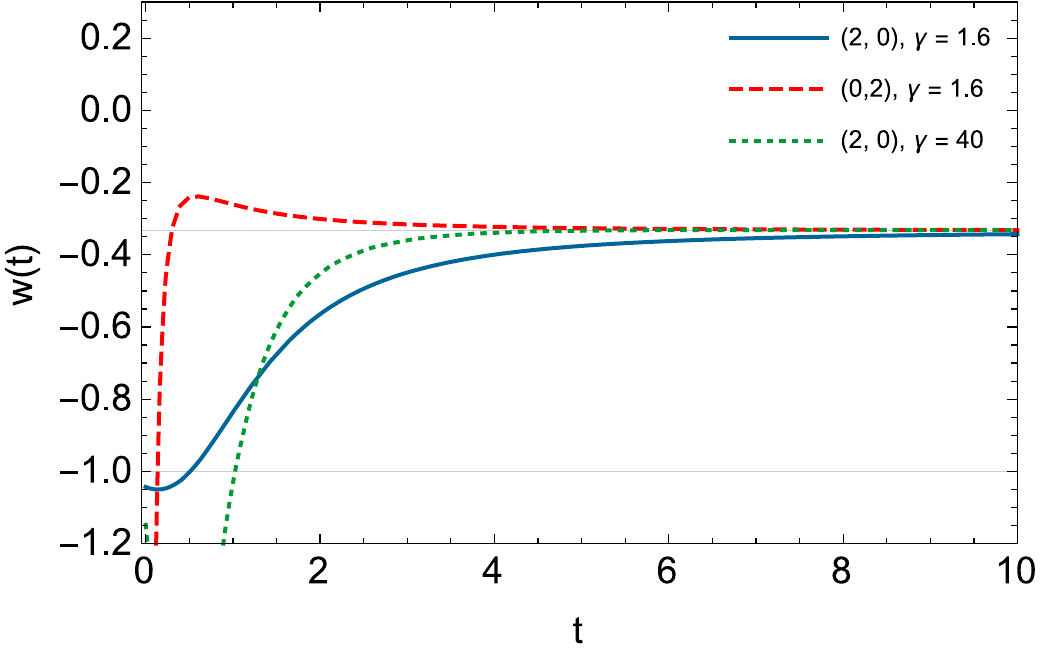}
  \endminipage
 \caption{\label{fig:w}  Behaviour of the equation of state parameter in comoving time for thermal initial states (top) and coherent initial states (bottom), for each value of $k = 1$ (left), $k = 0$ middle, $k= -1$ (right). In each case $w(t) \to -1/3$ regardless of the initial conditions or curvature. For late times $t > 5$ both the strong and weak energy conditions hold. Furthermore in all case  $w(t)$ approaches the asymptote faster for large $\gamma$.}
 \label{wvtkp}
\end{figure*}

An interesting feature of the model is that for large times, depicted in Fig.~\ref{wvtkp}, the system asymptotes to the relation $P(t)=-\frac{1}{3}\rho(t)$.     Although at early times the strong energy condition is generally (but not always) violated, at late times it is satisfied, with the equation of state settling down to the zero-acceleration case of $w=-1/3$.
This is a rather striking feature of our model that is robust to any changes in initial conditions as long as $\gamma >0$, and occurs for all values of $k$.  It is a consequence of the existence of the growing mode, which ensures
at late conformal times exponential growth of the scale factor, which translates into  asymptotic linear growth in comoving time.  From the Friedmann equations, if $w(t)>-1/3$ then the universe undergoes an decelerating expansion whereas if 
$w(t)<-1/3$ we have an accelerated expansion which is a necessary condition for inflating universes. A closer analysis of the behaviour of the different quantities involved in Eq. \eqref{eqofstate} shows that $\texttt{a}'(t)$ asymptotes to a constant finite value and, as shown in Fig.~\ref{a2vtkp}, the scale factor as a function of comoving time grows linearly with time. We thus  conclude that the asymptotic behavior of the equation of state is governed by the acceleration $ {\texttt{a}}''(t)$ which goes to zero for large times.   
This crucially depends on the positivity of $\gamma$, whose effects are most pronounced in the $k=1$ case.  In the limit $\gamma\to 0$, this case becomes purely oscillatory.  Increasingly large values of $\gamma$ both damp the oscillations and cause a more rapid growth in the scale factor, which asymptotes to a linear function of conformal time for all values of $k$. The fact that there are times for which the universe is expanding in an accelerated  fashion suggests that our model can be used as an alternative to inflationary models, but the complete investigation of this aspect is beyond the scope of the present work. 
  
\section{Conclusions} \label{sec.conclusion}

 We have explored the first implementation of CCG in a relativistic setting, showing how it can be implemented in an FRW spacetime.  We found that this yields  an alternative model for quantum cosmology, one in which the dynamical variables are quantum, and source a classical metric that influences test particles. By construction the evolution of the spacetime in the presence of such test particles is fundamentally non-unitary and results in an unavoidable decoherence of the quantum system and an arrow of time.  The non-unitarity is required in  order for a test particle to be influenced by the  scale factor in the CCG model. This results in an arrow of time and unavoidable decoherence of the quantum system.  
Furthermore, the big-bang singularity is removed, since the scale factor is now interpreted as the mean of a positive quantum variable  which is constrained by the uncertainty relations.

The net effect of this interaction is manifest in a form of time-dependent dark energy as our subsequent
investigation of  the evolution of the metric (as seen by an observer that measures the trajectories of  test particles) indicated. 
 For positive curvature $k>0$ we found that the cosmological evolution is generally characterized by oscillatory behaviour of the scale factor (consistent with the Friedmann solutions) that is eventually dominated by exponential growth in conformal time. Transforming to comoving coordinates, the equation of state parameter $w(t)$ initially undergoes oscillations that damp out, with this parameter reaching the asymptotic value of $-1/3$ at late times.  This condition, present for all values of $k$, is robust to initial conditions and is a consequence of the aforementioned exponential growth, which in turn is driven by the constraint of the decoherence rate of the quantum system. 
 
 We have thus shown than an observer in the universe will see the presence of a dark fluid as a consequence of test particles interacting with the metric. This form of fluid is characterized by $\gamma$, the fundamental parameter in our model. However the energy conditions are generically not satisfied at early times, with $|w(t)|>1$, although two notable exceptions are illustrated in the upper left of figure \ref{fig:w} (the solid and dashed curves) for thermal initial states. This suggests that a more sophisticated model could generically satisfy the energy conditions. Furthermore $\rho >0$ for $k\leq 0$. Of course the model we have presented here is overly simplistic, ignoring matter contributions and possible spatial inhomogeneities and anisotropies.  A more realistic cosmology must take such factors into account. 

We close by commenting on some special features of our model and future perspectives. From Eq. \eqref{eq:Uncond} we notice that the new parameter $\gamma$ has units of time. In cosmology the Hubble parameter $H$ gives the time scale of the universe, its age, and the size of the Hubble horizon. Our model introduces a new time scale that sets the scale of fluctuations of the different quantities and, as discussed in section \ref{sec.analysis}, these two scales are not completely independent. This parameter $\gamma$ is responsible (in this simplified version of the model) of the presence of a dark fluid making the universe expand. 

The model discussed in this work can be extended to consider perhaps more realistic scenarios. In particular we are interested in how a matter source interacting with the scale factor will modify the equations. In this scenario, there is no need to introduce the notion of a test particle to account for non unitary evolution. In fact, the CCG model states that in order for two quantum systems to interact gravitational (in this case scale factor and matter) both subsystems need to continuously have knowledge of the other subsystem properties, and this can be achieved by a classical communication channel or weak measurements. This (effective) weak measurements will break unitary evolution and  an observer in that universe that measure the matter field in order to describe its dynamics will induce decoherence on the joint system scale factor- matter therefore changing the dynamics as of a universe that behaves according to the Wheeler-deWitt equation.

Let us comment on the covariance properties of the CCG model as presented in this work. As formulated here, the model explicitly brakes unitarity evolution in the frame where the weak measurements performed by the test particles are held and therefore  the master equation for the scale factor was computed in the proper frame of the test particles. In this exploratory work we decided to work in conformal time, and thus both the test particles and the observers have the conformal time as their proper time. A more careful extension of the model should have this feature taken into account. For example, when matter is introduced one should look at its associated energy momentum tensor and in particular to its proper time. The proper time of the matter is then the frame in which the matter will interact with the metric and thus is in this frame where unitarity is broken. One should in principle write the master equation as a function of a the proper time of the matter. A similar description  that we presented in this work will therefore hold for an observer whose proper time is the proper time of matter. For any other observer, that does not share the same proper time as the matter, one will need to perform a change of reference to described the emergent dark fluid.   We postpone all this extensions to future work.

\section*{Acknowledgements}
The authors would like to thank Jake Taylor for his substantial contributions in motivating this work by proposing that dark energy may arise from in the classical channel formalism. His insight into the physical grounds for the particular measurement choice were helpful in understanding the model in a wider context. The authors are also thankful to Niayesh Afshordi and Romain Pascalie  for useful comments of the manuscript. This work was supported in part by the Natural Sciences and Engineering Research Council of Canada and The 
Australian Research Council grant CE110001013. P. C-U. gratefully acknowledges funding from CONACyT. N.A., P. C-U., and R.B.M. would like to thank the University of Queensland for their hospitality during the initial stages of this work. This research was supported in part by Perimeter Institute for Theoretical Physics. Research at Perimeter Institute is supported by the Government of Canada through the Department of Innovation, Science and Economic Development Canada and by the Province of Ontario through the Ministry of Research, 

\appendix

\section{Newtonian Classical Channel Model}\label{app.ccgmathematics}

In the following we show the difference between decoherence in unitary quantum mechanics and the CCG model.  We will use angle brackets $\langle \cdot\rangle$ to denote expectation value of a quantum observable, and $\mathcal{E}(\cdot)$ to denote average over the classical noise. Consider two massive particles innitially separated by a mean distance $d = \langle \hat{x}_1(0) - \hat{x}_2(0) \rangle$, interacting under a Newtonian gravitational potential. The potential can then be linearized about the mean separation, 
\begin{eqnarray}
\!\!\!\!-\frac{Gm_1m_2}{|\hat{x}_1 - \hat{x}_2|} &\approx& -\frac{Gm_1 m_2}{d}\left (1 - \frac{\delta \hat{x}_1 - \delta \hat{x}_2}{d}\right.  + \left. \frac{(\delta \hat{x}_1 - \delta \hat{x}_2)}{d^2} \right)\,,
\label{eq:}
\end{eqnarray}
where $\delta \hat{x}_i$ is the fluctuation about the mean seperation of the $i^{\mathrm{th}}$ particle and has zero mean. The cross term in the second order expansion is the first non-trivial quantum interaction between the two particles. Therefore, the lowest order Newtonian interaction is $H_I = K \hat{x}_1\hat{x}_2$ where $K = 2Gm_1m_2/d^3$,  using the notation from Ref.~\cite{2014NJPh...16f5020K}. Note that in general, the interaction Hamiltonian $H_I$, may result in entanglement between the separated particles. Working in the interaction picture and beginning with a separable, pure initial state $\hat{\rho}(0) = \hat{\rho}_1\otimes\hat{\rho}_2$, the joint system will unitarily evolve into 
\begin{eqnarray}
\hat{\rho}(\delta t) = e^{-iH_I \delta t/\hbar}\hat{\rho}_1\otimes\hat{\rho}_2 e^{-iH_I \delta t/\hbar}\,,
\label{eq:unitaryevolution}
\end{eqnarray}
after a time $\delta t$. The time $\delta t$ is assumed to be short enough such that the linearisation of $H_I$ is valid over the full duration. This is standard unitary evolution, and the joint system remains pure, $\mathrm{Tr}[(\hat{\rho}(\delta t))^2] = \mathrm{Tr}[(\hat{\rho}_1\otimes\hat{\rho}_2)^2]= 1$. However an observer who is unaware of particle two will see decoherence in the reduced state of particle one, $\hat{\rho}_1(\delta t) = \mathrm{Tr}_2[\hat{\rho}(\delta t)]$. In particular, note that even though the global evolution is non-dissipative, the observer sees decoherence in the description of their local quantum state. The decoherence is thus a consequence of thinking about the reduced evolution from the point of view of the observer, therefore necessarily requires the presence of an observer to make sense (see Fig.~\ref{fig:models} top-left).

In contrast, the CCG model postulates that the interaction $H_I$ is equivalent to a measurement and feedback process. Here we outline the relevant details presented in Ref.~\cite{2014NJPh...16f5020K} to highlight that the non-unitary dynamics in CCG is fundamental and \textit{independent} of the existence of any observer. In Newtonian CCG the interaction Hamiltonian is replaced by a feedback control Hamiltonian
\begin{eqnarray}
H_I = K \hat{x}_1 \hat{x}_2 \rightarrow H_{fb} = K\bar{x}_1\hat{x}_2  + K \bar{x}_2 \hat{x}_1\,,
\label{eq:hfb}
\end{eqnarray}
where $\bar{x}_i$ is the classical measurement outcome of a weak continuous measurement of $\hat{x}_i$. The measurement itself alters the unitary dynamics of the joint density matrix $(\hat{\rho})$ to the stochastic master equation, 
\begin{eqnarray}
d\hat{\rho}_c &=& -\frac{idt}{\hbar}[H, \hat{\rho}] - \frac{\Gamma_1dt}{2\hbar}[\hat{x}_2, [\hat{x}_1, \hat{\rho}]] - \frac{\Gamma_2dt}{2\hbar}[\hat{x}_2, [\hat{x}_2, \hat{\rho}]]  \nonumber \\
& &+ \sqrt{\frac{\Gamma_1}{\hbar}}dW_1\mathcal{H}[\hat{x}_1]\hat{\rho}_c + \sqrt{\frac{\Gamma_2}{\hbar}}dW_2\mathcal{H}[\hat{x}_2]\hat{\rho}_c\,,
\label{eq:meas1}
\end{eqnarray}
where $dW_i$ is a standard Wiener increment with $\mathcal{E}(dW_i) = 0$ and $\mathcal{E}(dW_idW_j) = dt \delta_{ij}$, and $\mathcal{H}[\hat{A}]\hat{\rho} = \hat{A}\hat{\rho} + \hat{\rho} \hat{A} - 2\langle \hat{A}\rangle$ for any operator $\hat{A}$. The joint state of the system is conditioned (subscript $c$) on the knowledge of the measurement outcome 
\begin{equation}\label{x-bar}
\bar{x}_i = \langle \hat{x}_i \rangle_c +  \sqrt{\hbar / 2 \Gamma_i}dW_i/dt\,,
\end{equation}
 and $\Gamma_i$ describes the strength of the measurement. While the derivative $dW/dt$ is not formally defined, it can be understood as a white noise process, $dW/dt = \xi(t)$ where $\mathcal{E}[\xi(t) \xi(t')] = \delta(t - t').$ This modification from unitary dynamics is from the postulate that gravity is mediated by a classical information channel and has noting to do with the existence of an observer describing a reduced quantum state. After the instantaneous weak measurement is made, the joint system evolves under a unitary generated by feedback Hamiltonian \eqref{eq:hfb}, $U_{fb} = \exp{(-i dt H_{fb}/\hbar)}$, i.e. 
\begin{eqnarray}
\hat{\rho}(t + dt)_c = U_{fb} [\hat{\rho}(t) + d\hat{\rho}_c] U_{fb}^\dagger\,,
\end{eqnarray}
and the systematic interaction $H_I$ is recovered in the unitary part of the evolution. The non-unitary components of Eq.~\eqref{eq:meas1}, along with the noise in the feedback unitary (i.e. the $dW$ term in $H_{fb} dt$) result in decoherence in the joint quantum state when averaging over all possible outcomes (equivalent to an observer making an ensemble average of all possible measurement outcomes, or simply being unaware that the measurement happened).

At this point we diverge slightly from the discussion in Ref.~\cite{2014NJPh...16f5020K}, and instead of treating each particle symmetrically, we consider $\hat{x}_1$ as a `source' particle and $\hat{x}_2$ as a `test' particle, although the distinction is made arbitrarily. Since we now have a `test' particle, we are not concerned with the back reaction from the test onto the source, and therefore the effective feedback Hamiltonian to consider is $H_{fb} = K\bar{x}_1\hat{x}_2$ to generate the dynamics of the test particle. In this case $H_{fb}$ does not affect the Hilbert space of the source, and the measurement of the source (required for $H_{fb}$), does not affect the Hilbert space of the test particle. Therefore if the joint state is initially separable, $\hat{\rho}_0 = \hat{\rho}_1\otimes \hat{\rho}_2$, and there is no other interaction present, then the joint state will remain separable at all times, $\hat{\rho}(t) = \hat{\rho}_1(t)\otimes \hat{\rho}_2(t)$. However, the introduction of the measurement of the source implies fundamental decoherence in its quantum state and is required by the simple existence of the test particle (see Fig.~\ref{fig:models} top-right). In this asymmetric treatment between the two particles, there is no way to minimize the decoherence rate, but we can conclude that there must be a non-zero decoherence rate, $\Gamma_1>0$, of the source particle to determine the dynamics of the test particle. 

This one sided description is analogous to what we consider in the cosmological case. The dynamics of the test particle depends on classical information from the scale factor state. We therefore suggest, that in CCG there is some fundamental, observer independent decoherence. Since we do not consider the back reaction from the presence of a test particle on the scale factor the decoherence rate cannot be minimized, and is left as a free, but strictly positive parameter.  Further exploration should include the back reaction and thus introduce a noise minimization procedure.

\section{Master Equation}\label{app.master}

Following~\ref{app.ccgmathematics} we derive the master equation for the cosmological system.  We propose that the state is subject to weak continuous measurement of the variable $a^2$\footnote{This is because (classically) it is exactly the factor $a^2$ that appears in the metric function, and therefore the trajectory of any test particle explicitly can only depend on $a^2$.}. The way a test particle responds to the influence of the quantum scale factor through a classical metric function is via the result of a weak measurement. In this case the metric function is given by, $ds^2 = \bar{a}^2(-dt^2+dx^2)$ where $\bar{a}^2 = \langle \hat{a}^2\rangle_c + \sqrt{\frac{\hbar}{\gamma}}dW/dt$, relabeling $\Gamma\to \gamma$ from
\eqref{x-bar}.
This is analogous to the way a test particle responds to the Newtonian potential though $H_{fb} = K\bar{x}_1\hat{x}_2$. The presence of the weak measurement on the quantum scale factor changes the evolution of the quantum state $\hat{\rho}$, resulting in the stochastic master equation \cite{Altamirano:2016knc,2016arXiv160504302G}
\begin{eqnarray}
d\hat{\rho}_c &=& -\frac{i}{\hbar}[\hat{H}, \hat{\rho}]d\tau - \frac{\gamma}{8\hbar} [\hat{a}^2, [\hat{a}^2, \hat{\rho}]d\tau 
 - \sqrt{2\hbar}dW\mathcal{H}[\hat{a}^2]\hat{\rho}_c\,,
\label{eq:b2}
\end{eqnarray}
where we have assumed a continuous Gaussian measurement of $\hat{a}^2$, and the subscript $c$ refers to fact that the change in $\hat{\rho}$ is conditioned on the measurement result $\bar{a}^2$. Any observer who is unaware of the measurement outcome $\bar{a}^2$, will describe the state as an ensemble average over the measurement process, $d\hat{\rho}_c\rightarrow \mathcal{E}(d\hat{\rho}_c) = d\hat{\rho} $, with the corresponding ensemble averaged metric, $\bar{a}^2 \rightarrow \mathcal{E}(\bar{a}^2) = \langle \hat{a}^2\rangle \equiv \texttt{a}^2$ . Consequently the corresponding spacetime  is given by
\begin{eqnarray}
ds^2 = \langle \hat{a}^2(\tau) \rangle (-dt^2+dx^2) = \texttt{a}^2(\tau) (-dt^2 +dx^2)\,,
\label{eq:q-metric}
\end{eqnarray}
where the evolution of $\langle \hat{a}^2\rangle$ is given by Eq.~\eqref{eq:b2} using $\langle \dot{\h A}\rangle=\mathrm{Tr}[\h A\dot{\hat{\rho}}]$ for any operator $\h A$.

The condition $\gamma \neq 0$ is required in order for the test particle to feel the presence of the scale factor in the classical metric function, and $\gamma \geq 0$ is required preserve the positivity of $\hat{\rho}$~\cite{lindblad_generators_1976,wiseman2009quantum}. Therefore $\gamma >0$ is a requirement for the model to be physical. Fluctuations in the measurement record (of order $dW \sqrt{\hbar/ \gamma}$) induce fluctuations in the metric function. However, any observer making multiple measurements would average over all fluctuations, and only see the ensemble averaged dynamics~\cite{downes_optimal_2011}, i.e. Eq.~\eqref{eq:b2} for the quantum system.  The equations of motion for the first and second order moments are found from equation~\eqref{eq:b2}  to be 
\begin{eqnarray}
~~~~~~~~\frac{d}{d\tau}{\langle \hat{a} \rangle}&=&-\langle \hat{\pi}\rangle/2\,, \\
~~~~~~~~\frac{d}{d\tau}{\langle {\hat{\pi}} \rangle} &=&2k \langle \hat{a}\rangle\,, \\
~~~~~~~~\frac{d}{d\tau}{\langle \hat{a}^2 \rangle}&=& -\langle \hat{a}\hat{\pi}+\hat{\pi}\hat{a}\rangle/2\,, \label{a2}\\
~~~~~~~~\frac{d}{d\tau}{\langle \hat{\pi}^2 \rangle}&=& 2k\langle \hat{a}\hat{\pi}+\hat{\pi}\hat{a}\rangle + \gamma \langle \hat{a}^2 \rangle\,, \label{pi2}\\
\frac{d}{d\tau}{\langle \hat{a}\hat{\pi}+\hat{\pi}\hat{a}\rangle}&=& -\langle\hat{\pi}^2\rangle +4k\langle \hat{a}^2\rangle\,. \label{comb2}
\end{eqnarray}
where we have assumed the Hamiltonian is given by the canonical Hamiltonian for an empty FRW space~\ref{eq:hamilt}. Note that setting $\gamma = 0$ the standard Friedmann solutions are recovered. The modification $\gamma > 0 $ in the CCG model is a result of postulating that a test particle responds to the scale factor though a classical metric function, avoiding the complications of considering a quantized manifold. The decoherence is \textit{only} on the scale factor - this is because we have only considered the presence of `test' particles in the universe, similar to the source-test description in the Newtonian case in section~\ref{app.ccgmathematics}. Note that our approach is equivalent to the scale factor (and more generally, all metric degrees of freedom) repeatedly interacting with additional matter degrees of freedom whose net effect is to repeatedly `measure' the scale factor. These additional degrees of freedom may be thought of as quantum ``test particles" in the spacetime that naturally measure the scale factor along their trajectory.

\section{Solutions to the evolution equations}

The coupled system of differential equations \eqref{A2}--\eqref{Comb2} can be straightforwardly solved to find $\langle \hat{a}^2 \rangle(\tau)$, and thus the resulting spacetime. This system can be written on the form
 $\frac{d \vec{x}(\tau)}{d\tau}=A\vec{x}(\tau)$, with $\vec{x}(\tau) = ( \langle{\hat{a}^2}\rangle, \langle {\hat{\pi}^2}\rangle, \langle {\hat{a}\hat{\pi}+\hat{\pi}\hat{a}}\rangle)^{\textrm{T}}$ and  

\begin{eqnarray}\label{3.1}
A= \left(
\begin{array}{lll}
    0 & 0 & -\frac{1}{2} \\\\
    \gamma & 0 & 2k\\\\
    4k & -1 & 0
\end{array}\right) \,.
\end{eqnarray}

 Assuming a solution $\vec{x}(t)=\vec{\eta}e^{\lambda \tau}$, then $\vec{\eta}$ and $\lambda$ are the eigenvectors and eigenvalues of $A$ respectively where the eigenvalues are solutions to the characteristic equation 
\begin{equation}\label{characteristic}
\lambda^3+ 4 k \lambda-4\ \frac{\gamma}{8}=0\,,
\end{equation}
yielding
 \begin{eqnarray}\label{lam-m}
\lambda_m &=& \frac{\varsigma^2 e^{2m\pi i/3} - 12 k e^{-2m\pi i/3}}{3\varsigma}\,, \nonumber\\
 \varsigma &=& 3\left(\frac{\gamma}{4} +2\sqrt{\Delta}\right)^{1/3}\,,
\end{eqnarray}
where $m=1,2,3$. 
It is straightforward to show that the characteristic equation always has one positive real root if $\gamma>0$, which must be the case for  physically sensible measurements. The general nature of the solutions is determined by the sign of $\Delta$, where
 \begin{eqnarray}
   \Delta = \frac{16}{27}k^3+ \left (\frac{\gamma}{8}\right)^2,
 \end{eqnarray}
yielding distinct real roots ( $\Delta < 0$), multiple real roots ($\Delta=0$), or one real and two complex conjugate roots ($\Delta > 0$).  The former two cases occur only for $k< 0$;  clearly  $\Delta < 0$  if and only if $(\gamma/8)^2< \frac{16 |k|^{3}}{27}$ for $k<0$.  The eigenvectors are $\vec{\eta_i}= (-\frac{1}{2 \lambda_i} ,  \frac{-2k+\lambda_i^2}{\lambda_i}, 1)^{\textrm{T}}$ where the $\lambda_i$ are solutions to   \eqref{characteristic}.

We find there is  always one positive real solution to eq. \eqref{characteristic}. This means that there is always one exponentially growing mode, which makes the universe expand. Furthermore the real parts of the other two roots are always negative, and so the other modes will exponentially decay ($\Delta \leq 0$), or oscillate with an exponentially decaying envelope ($\Delta> 0$).  Since a general solution will be a linear combination of the eigen-solutions
\begin{equation}\label{lin-comb}
\vec{x}(t)=\sum_i c_i\vec{\eta_i}e^{\lambda_i \tau}\,,
\end{equation}
the solution will in general asymptote to one that grows exponentially with time. We note that if  initial conditions are chosen such that the coefficient $c_+$ of the positive real root $\lambda_+$ vanishes then $\langle \hat{a}^2\rangle$  and $\langle\hat{\pi}^2\rangle$ will become arbitrarily small, violating the Heisenberg uncertainty principle.  All valid initial quantum states must have $c_+$ nonzero.

\section*{References}
\bibliography{Databaze}

\begin{thebibliography}{10}

\bibitem{baym_two-slit_2009}
Gordon Baym and Tomoki Ozawa.
\newblock Two-slit diffraction with highly charged particles: {Niels} {Bohr}'s
  consistency argument that the electromagnetic field must be quantized.
\newblock {\em PNAS}, 106(9):3035--3040, March 2009.

\bibitem{Polchinski:1998rq}
J.~Polchinski.
\newblock {\em {String theory. Vol. 1: An introduction to the bosonic string}}.
\newblock Cambridge University Press, 2007.

\bibitem{Rovelli:2010bf}
Carlo Rovelli.
\newblock {Loop quantum gravity: the first twenty five years}.
\newblock {\em Class. Quant. Grav.}, 28:153002, 2011.

\bibitem{modesto_super-renormalizable_2012}
Leonardo Modesto.
\newblock Super-renormalizable quantum gravity.
\newblock {\em Phys. Rev. D}, 86(4):044005, August 2012.

\bibitem{stelle_renormalization_1977}
K.~S. Stelle.
\newblock Renormalization of higher-derivative quantum gravity.
\newblock {\em Phys. Rev. D}, 16(4):953--969, August 1977.

\bibitem{moffat_finite_1990}
J.~W. Moffat.
\newblock Finite nonlocal gauge field theory.
\newblock {\em Phys. Rev. D}, 41(4):1177--1184, February 1990.

\bibitem{cornish_quantum_1992}
N.j. Cornish.
\newblock Quantum non-local gravity.
\newblock {\em Mod. Phys. Lett. A}, 07(07):631--639, March 1992.

\bibitem{kiefer_conceptual_2013}
Claus Kiefer.
\newblock Conceptual {Problems} in {Quantum} {Gravity} and {Quantum}
  {Cosmology}.
\newblock {\em ISRN Mathematical Physics}, 2013:1--17, 2013.

\bibitem{carlip_is_2008}
S.~Carlip.
\newblock Is quantum gravity necessary?
\newblock {\em Class. Quantum Grav.}, 25(15):154010, 2008.

\bibitem{albers_measurement_2008}
Mark Albers, Claus Kiefer, and Marcel Reginatto.
\newblock Measurement analysis and quantum gravity.
\newblock {\em Phys. Rev. D}, 78(6):064051, September 2008.

\bibitem{boughn_nonquantum_2009}
Stephen Boughn.
\newblock Nonquantum {Gravity}.
\newblock {\em Found Phys}, 39(4):331--351, February 2009.

\bibitem{peres_hybrid_2001}
Asher Peres and Daniel~R. Terno.
\newblock Hybrid classical-quantum dynamics.
\newblock {\em Phys. Rev. A}, 63(2):022101, January 2001.

\bibitem{Ahmadzadegan:2016wsm}
Aida Ahmadzadegan, Robert~B. Mann, and Daniel~R. Terno.
\newblock {Classicality of a quantum oscillator}.
\newblock {\em Phys. Rev.}, A93(3):032122, 2016.

\bibitem{diosi_models_1989}
L.~Di{\'o}si.
\newblock Models for universal reduction of macroscopic quantum fluctuations.
\newblock {\em Phys. Rev. A}, 40(3):1165--1174, August 1989.

\bibitem{penrose_gravitys_1996}
Roger Penrose.
\newblock On {Gravity}'s role in {Quantum} {State} {Reduction}.
\newblock {\em Gen Relat Gravit}, 28(5):581--600, May 1996.

\bibitem{2014NJPh...16f5020K}
D~Kafri, J~M Taylor, and G~J Milburn.
\newblock A classical channel model for gravitational decoherence.
\newblock {\em New Journal of Physics}, 16(6):065020, 2014.

\bibitem{nielsen_conditions_1999}
M.~A. Nielsen.
\newblock Conditions for a {Class} of {Entanglement} {Transformations}.
\newblock {\em Phys. Rev. Lett.}, 83(2):436--439, July 1999.

\bibitem{aharonov_how_1988}
Yakir Aharonov, David~Z. Albert, and Lev Vaidman.
\newblock How the result of a measurement of a component of the spin of a spin-
  {\textbackslash}textit\{1/2\} particle can turn out to be 100.
\newblock {\em Phys. Rev. Lett.}, 60(14):1351--1354, April 1988.

\bibitem{gardiner_input_1985}
C.~W. Gardiner and M.~J. Collett.
\newblock Input and output in damped quantum systems: {Quantum} stochastic
  differential equations and the master equation.
\newblock {\em Phys. Rev. A}, 31(6):3761--3774, June 1985.

\bibitem{diosi_quantum_1995}
Lajos Diosi.
\newblock Quantum dynamics with two {Planck} constants and the semiclassical
  limit.
\newblock {\em arXiv:quant-ph/9503023}, March 1995.

\bibitem{diosi_quantum_2000}
Lajos Di{\'o}si, Nicolas Gisin, and Walter~T. Strunz.
\newblock Quantum approach to coupling classical and quantum dynamics.
\newblock {\em Phys. Rev. A}, 61(2):022108, January 2000.

\bibitem{kafri_bounds_2015}
D.~Kafri, G.~J. Milburn, and J.~M. Taylor.
\newblock Bounds on quantum communication via {Newtonian} gravity.
\newblock {\em New J. Phys.}, 17(1):015006, 2015.

\bibitem{khosla_detecting_2016}
Kiran Khosla and Natacha Altamirano.
\newblock Detecting gravitational decoherence with clocks: {Limits} on temporal
  resolution from a classical channel model of gravity.
\newblock {\em arXiv:1611.09919}, November 2016.

\bibitem{Altamirano:2016fas}
Natacha Altamirano, Paulina Corona-Ugalde, Robert~B. Mann, and Magdalena Zych.
\newblock {Gravity is not a Pairwise Local Classical Channel}.
\newblock 2016.

\bibitem{jacobs_straightforward_2006}
Kurt Jacobs and Daniel~A. Steck.
\newblock A straightforward introduction to continuous quantum measurement.
\newblock {\em Contemporary Physics}, 47(5):279--303, September 2006.

\bibitem{downes_optimal_2011}
T.~G. Downes, G.~J. Milburn, and C.~M. Caves.
\newblock Optimal {Quantum} {Estimation} for {Gravitation}.
\newblock {\em arXiv:1108.5220}, August 2011.

\bibitem{kar_maxwell_1993}
G.~Kar, M.~Sinha, and S.~Roy.
\newblock Maxwell equations, nonzero photon mass, and conformal metric
  fluctuation.
\newblock {\em Int J Theor Phys}, 32(4):593--607, April 1993.

\bibitem{Wiltshire}
D.L. Wiltshire.
\newblock An introduction to quantum cosmology.
\newblock {\em arXiv:gr-qc/0101003v2}, 2010.

\bibitem{MONIZQC}
P.V. MONIZ.
\newblock Supersymmetric quantum cosmology shaken, not stirred.
\newblock {\em International Journal of Modern Physics A}, 11(24):4321--4382,
  1996.

\bibitem{Calcagni10}
Gianluca Calcagni.
\newblock Fractal universe and quantum gravity.
\newblock {\em Phys. Rev. Lett.}, 104:251301, June 2010.

\bibitem{Maartens2010}
Roy Maartens and Kazuya Koyama.
\newblock Brane-world gravity.
\newblock {\em Living Reviews in Relativity}, 13(1):5, 2010.

\bibitem{tachyon}
John McGreevy and Eva Silverstein.
\newblock The tachyon at the end of the universe.
\newblock {\em JHEP}, 0508:090, 2005.

\bibitem{lindblad_generators_1976}
Goran Lindblad.
\newblock On the generators of quantum dynamical semigroups.
\newblock {\em Comm. Math. Phys.}, 48:119--130, 1976.

\bibitem{wiseman2009quantum}
Howard~M Wiseman and Gerard~J Milburn.
\newblock {\em Quantum measurement and control}.
\newblock Cambridge University Press, 2009.

\bibitem{bassi_models_2013}
Angelo Bassi, Kinjalk Lochan, Seema Satin, Tejinder~P. Singh, and Hendrik
  Ulbricht.
\newblock Models of wave-function collapse, underlying theories, and
  experimental tests.
\newblock {\em Rev. Mod. Phys.}, 85(2):471--527, April 2013.

\bibitem{PhysRevLett.118.021102}
Thibaut Josset, Alejandro Perez, and Daniel Sudarsky.
\newblock Dark energy from violation of energy conservation.
\newblock {\em Phys. Rev. Lett.}, 118:021102, Jan 2017.

\bibitem{Altamirano:2016knc}
Natacha Altamirano, Paulina Corona-Ugalde, Robert~B Mann, and Magdalena Zych.
\newblock Unitarity, feedback, interactions, dynamics emergent from repeated
  measurements.
\newblock {\em New Journal of Physics}, 19(1):013035, 2017.

\bibitem{2016arXiv160504302G}
Daniel Grimmer, David Layden, Robert~B. Mann, and Eduardo
  Mart\'{\i}n-Mart\'{\i}nez.
\newblock Open dynamics under rapid repeated interaction.
\newblock {\em Phys. Rev. A}, 94:032126, Sep 2016.

\end{thebibliography}
\bibliographystyle{unsrt}

\end{document}